\documentstyle[preprint,aps]{revtex}
\tightenlines

\newcommand {\be}{\begin{equation}}
\newcommand {\ee}{\end{equation}}
\newcommand{\bey}{\begin{eqnarray}}
\newcommand{\eey}{\end{eqnarray}}

\begin{document}
\title{Finite-size corrections in Lyapunov Spectra for Band Random Matrices}
\author {T. Kottos$^{1,2}$, A. Politi$^{3}$, F.M. Izrailev$^{4,5}$ \\
$^1$ Department of Physics of Complex Systems, The Weizmann Institute of
Science, Rehovot 76100, Israel \\
$^2$ Department of Physics, University of Crete and Research Centre of Crete,
PO Box 1527, 71110 Heraklion, Crete, Greece.\\
$^3$ Istituto Nazionale di Ottica 50125 Firenze, Italy and \\
Istituto Nazionale di Fisica Nucleare, Sezione di Firenze\\
$^4$ Instituto de Fisica, Universidad Autonoma de Puebla, \\
Apdo. Postal J-48, Col. San Manuel, Puebla, Pue. 72570, Mexico\\
$^5$ Budker Institute of Nuclear Physics, Novosibirsk 630090, Russia
}
\date{\today}
\maketitle

\begin{abstract}
\begin{center}
\parbox{14cm}
{The transfer matrix method is applied to quasi one-dimensional
and one-dimensional disordered systems with long-range interactions
described by band random matrices. We investigate the convergence properties
of the whole Lyapunov spectra of finite samples as a function of the
bandwidth and of the sample length. Two different scaling laws are found 
at the maximal and minimal Lyapunov exponents.}
\end{center}
\end{abstract}

\pacs{
PACS numbers: 71.55J, 05.45.+b}


\newpage
\section{\bf Introduction}

Lyapunov exponents represent one of the main tools in the study of both
disordered and dynamical systems. In the former case, they arise from the
application of the transfer matrix method and allow determining, e.g., the
localization length, which corresponds to the inverse of the minimum
positive Lyapunov exponent. In the latter case, starting from the evolution
in the tangent space, one is interested in determining, among other
quantities, the Kolmogorov-Sinai entropy that can be expressed as the sum of
the positive exponents.

Recently, a growing attention has been devoted to the study
of high-dimensional systems such as coupled maps, chains of nonlinear
oscillators, dynamical models with delayed feedback, disordered systems in
two and three dimensions, and 1-dimensional models with long range
interaction. In all these
cases, the so-called Lyapunov spectrum is defined as the sequence of
the Lyapunov exponents $\gamma_i$ (ordered for increasing/decreasing 
size) represented as a function of $i/D$, where $D$ is the 
total number of exponents.
Many numerical simulations and analytical arguments 
indicate the existence of a limit spectrum for $D \to \infty$ (see, e.g.,
\cite{LPT96} and references therein). In contrast
to the largest and smallest exponents, for which some rigorous mathematical 
results have been obtained, the properties of the ``bulk'' of the Lyapunov 
spectrum are less understood. 

In Ref.~\cite{KPIR96}, the scaling properties of the Lyapunov spectrum have
been studied in disordered systems described by infinite Band Random
Matrices (BRM). Such matrices have been extensively investigated in
connection with 1-dimensional Anderson-type models with long range random
hopping as well as with quasi 1-dimensional thin wires (see e.g.
\cite{theory} and references
therein). In particular, the scaling properties of the eigenfunction 
localization lengths proved to be in accordance with the predictions
\cite{CGIS90,CMI90} based on results for dynamical quantum models that are
strongly chaotic in the classical limit.

Since any disordered sample used in practical applications is finite,
it is useful not only to study the asymptotic value of the Lyapunov
exponents but also the so-called effective exponents, i.e. the exponents
actually observed in samples of finite length $N$. Moreover, information
about the statistical properties of effective Lyapunov exponents can
shed further light on, e.g., the fluctuations of the conductance in
the metallic regime \cite{M96,P84,I86,PZIS90,KIPC97}.

In this paper, we investigate the convergence properties of the effective
Lyapunov exponents of finite BRMs associated with finite samples embedded in
an otherwise perfectly ordered lattice. Our investigation suggests that
different parts of the spectrum exhibit different convergence properties.
This is particularly clear from our data for the maximal Lyapunov exponent
as compared to the bulk of the spectrum.

The outline of the paper is as follows. In Sec.~2 the connection between
the Hamiltonian band matrix model and the conductance of disordered samples
is summarized with the purpose of introducing the appropriate Lyapunov
exponents. In Sec.~3, we recall some known results of scaling theory in
similar problems: they will represent the starting point for
the numerical investigation carried on in the following Sec.~4.
Some final comments and conclusions are presented in
the last Sec.~5.

\section{\bf The Model}

The general model describing quasi 1-dimensional or 1-dimensional systems
with long range hopping is given by the Schr\"odinger equation
\begin{equation}  
\label{eqmo}
i\,\frac{dc_n(t)}{dt}=\,\sum_{m=n-b}^{n+b}H_{n,m}c_m (t) \quad ,
\end{equation}
where $c_n(t)$ is the probability amplitude for an electron to be at site $n$
and $H_{n,m}$ is a symmetric band random matrix. Specifically, the entries 
of $ H_{n,m} $ are independent Gaussian variables with zero mean and
variance $\sigma^2 = 1 + \delta_{n,m}$ for $|n-m| \leq b$, while
the matrix elements outside the band are all set equal zero.
In 1-dimensional geometry, the parameter $b$ defines the hopping range
between neighboring sites, while in the quasi 1-dimensional interpretation,
this parameter has the meaning of the number of transverse channels for the
scattering waves along a thin wire \cite{theory}.

The insertion of a disordered sample of length
$N$ into a perfectly ordered lattice requires the definition of two proper
leads at the extrema of the sample. At variance with the standard
Anderson model, where only nearest-neighbor couplings are present, the
long-range hopping terms in our model allow some freedom in the
structure of the ordered leads. A reasonable choice consists in assuming a
band structure in the ordered lattice with the same width $b$ and the
hopping elements $H_{n,m}$ set all equal to $U$ (for the sake of simplicity
we choose $U=1$), while the hopping amplitudes coupling the leads with
sample-sites are randomly chosen with the same distribution as in the core
of the sample (the intermediate regions connecting the samples with the
leads will be hereafter called ``contacts'': they extend over $b$ sites).
As an example, we show below the Hamiltonian structure for $b=2$
(asterisks mark random elements)
\[
\matrix {
\ldots &{\ldots} &1 & 1& {}& {}    & {}& {}& {}& {}& {}& {}\cr
1      &1 &1 & 1& * & {}    & {}& {}& {}& {}& {}& {}\cr
{}     &1 &1 & 1& * & *     & {}& {}& {}& {}& {}&  {}\cr
{}     &{}&* & *& * & *     & \ldots & {}&  {}&  {}& {}& {}\cr
{}     &{}&{}&{}&{}     &    \ldots &  *&  *&  *& * & {}& {}\cr
{}     &{}&{}&{}&{} &   {}&  *&  *&  1&  1& 1& {}\cr
{}     &{}&{}&{}&{} &  {}&  {}&  *&  1&  1&  1& \ldots\cr
}
\]

The eigenvalue equation is obtained by inserting the standard Ansatz
$c_n(t)=\exp (-iEt) \psi_n$ in Eq.~(\ref{eqmo}). The resulting equation can
be casted in the form of a $2b$-dimensional linear map along the spatial
direction, 
\begin{equation}  
\label{eq:def1}
\Psi(n+1) = T(n) \Psi(n) \; ,
\end{equation}
where $\Psi_i(n) \equiv \psi_{n+b-i}$ and the matrix $T(n)$ is defined as
follows,
\begin{eqnarray}  \label{eq:mmap}
&& [T(n)]_{1,j} = \frac{1}{H_{n,n+b}} \left( \delta_{j,b}E - H_{n,n+b-j}
\right)  \nonumber \\
&& [T(n)]_{i,j} = \delta_{i-1,j}; \quad \quad [T(n)]_{i,2b}=0; \quad\quad 2
\le i \le 2b
\end{eqnarray}
In this picture, an eigenstate of Eq.~(\ref{eqmo}) can be treated as a
``trajectory'' of the random map (\ref{eq:mmap}) and its localization
properties are determined by the Lyapunov exponents.

In a previous paper \cite{KPIR96} we investigated the shape of the Lyapunov
spectrum in the limit of infinitely extended disordered samples. Here, since
we are interested in finite samples of size $N$, one should introduce the
transfer matrix $T= \prod_{n=1}^{N} T(n)$, which couples two opposite leads.
As was shown in \cite{M96}, the matrix $T$ satisfies the following property,
not shared by the single matrix $T(n)$,
\begin{equation}  
\label{symp1}
T^t \Sigma T =\Sigma \quad , \quad\quad
\Sigma= \pmatrix {0 & S\cr -S^t & 0\cr }
\end{equation}
where $S$ is a lower triangular matrix of size $b$, with $S_{ij}=1$,
$i\ge j$. In fact, this property corresponds to the flux
conservation in the scattering process of a wave through the sample.

It is convenient to describe the scattering states in terms of
eigenfunctions of the free dynamics occurring in the ordered region. The
eigenvalues of the corresponding Hamiltonian, defined by setting all random
elements equal to 1, are
\begin{equation}
\label{free2}
E(p)=1+2\cos p+\ldots 2\cos (bp)={\frac{{\sin [(2b+1)(p/2)]}}{{\sin (p/2)}}}
,\quad \quad p\in (-\pi ,\pi )  
\end{equation}
while the corresponding eigenvectors are
\begin{equation}
\label{free1}
\psi _n(p)={\frac 1{{\sqrt{2\pi }}}}e^{\pm inp} \;.
\end{equation}
For any fixed energy value $\tilde{E}$, there are $\nu \le b$ pairs of
opposite real solutions of the equation $E(p_k)=\tilde{E}$. Each pair
corresponds to an open channel, or propagation mode, sustaining waves with
opposite velocities. In this paper, we limit ourselves to study the case
$E=0$, when all channels are open, i.e. $\nu =b$, and the admissible momenta
are equispaced. In what follows we shall assume that the momenta $p_k$ are
ordered in such a way that positive velocities correspond to the first
$b$ elements,
\begin{equation}
\label{vel1}
p_1,\,p_2,\,\ldots p_b,\,-p_1,\,\ldots -p_b\;,\;p_k=(-1)^k{\frac{{2\pi k}}
{{2b+1}}}\;,\;k=1,\ldots ,b\;.
\end{equation}
In fact, one can see that the corresponding velocities $v(p)=dE/dp$ for
$E=0$ are given by
\begin{equation}
v_k={\frac{2b+1}{2\sin[{\frac{\pi k}{2b+1}}]}}\;,\;v_{k+b}=-v_k\;,\;k=1,%
\ldots ,b\;.  \label{vel}
\end{equation}
Instead of defining the initial state in the scattering process in terms of
the probability amplitude in $2b$ consecutive sites (as required by
the standard representation of the vector $\Psi (n)$), one can refer to the
$2b$ amplitudes of the plane waves sustained by the ordered lattice. It can
be easily checked that the transformation to pass from the momentum to the
position representation is defined by the matrix
\begin{equation}
\lbrack U(n)]_{j,k}=\exp \bigg( i(j+n)p_k\bigg) \;,
\end{equation}
so that the scattering matrix can be written as
\begin{equation}
\label{tran3}
M=Z^{-N}U(0)^{-1}TU(0)\;
\end{equation}
where $Z$ is a diagonal matrix the entries of which, $Z_{j,j} = e^{ip_j}$,
account for the phase difference between the sites $n=1$ and $n=N$.
The matrix $M$ has an almost symplectic structure; indeed, it satisfies the
relation $M^{\dagger}VM=V$, where $V$ is a diagonal matrix with
$V_{j,j} = v_j$ in the same order as before \cite{M96}.

For the determination of the conductance, one needs to introduce the matrix $F$
connecting flux amplitudes, 
\begin{equation}  \label{trvel}
F=\Gamma M \Gamma^{-1} \; ,
\end{equation}
where the diagonal $2b \times 2b$ matrix $\Gamma$ is defined as
$\Gamma_{i,j} = \delta_{i,j} \sqrt {|v_i|}$. The above transformation
is equivalent to the normalization of the scattering matrix and it
takes into account that the waves propagate with different velocities in
different open channels.

The transfer matrix $F$ connecting the incoming with the outgoing flux
amplitudes in the various channels turns out to be symplectic as it
satisfies the relation
\begin{equation}
\label{flux}
F^{\dagger} \sigma_3 F = \sigma_3 \quad ,
\end{equation}
where $\sigma_3$ is a generalized Pauli $\sigma_z$ matrix,
\begin{equation}
\sigma_3 = \pmatrix{ 1 & 0 \cr 0 & -1 \cr} \; ,
\end{equation}
and 1 denotes a $b\times b$ identity matrix. Notice that condition
(\ref{flux}) follows essentially from the flux conservation, i.e. from the
unitarity of the scattering matrix.

The matrix $F$ is the key ingredient for the determination of the
conductivity from the Landauer formula (see, e.g., \cite{P84} for a general
derivation and \cite{KIPC97} for an application in the specific case of BRMs).
More precisely, it is necessary to compute the Lyapunov exponents
\be
  \gamma_i(N) =  \frac{1}{2} \ln m_i(N) \quad ,
\ee
where $m_i(N)$ denote the (real) eigenvalues of the matrix $F^{\dagger }F$.
The Lyapunov exponents will be conventionally ordered from the maximum to
the minimum one, namely
$\gamma_1(N)>\gamma _2(N)>\ldots >\gamma _i(N)>\ldots > \gamma _{2b}(N)$.
Because of the symplectic structure of $F$, the exponents are arranged in
$b$ pairs with opposite values
($\gamma _i(\nu )=-\gamma _{2b-i+1}(\nu )$); for this reason, in the
following, we will always report only the positive exponents (i.e. $i \le b$).

In the limit of infinitely long samples ($N \to \infty$), the ergodic
multiplicative theorem \cite{O68} ensures that the statistical fluctuations
of the effective Lyapunov exponents vanish: in fact, the quantities
$\gamma_i(\infty)$'s are, by definition, the Lyapunov exponents of the
infinite product of matrices composing $F$. Moreover, in the
limit $N \to \infty$, the effect of the similarity transformations
involved in the definition (\ref{tran3},\ref{trvel}) of $F$ becomes
negligible, so that the values of $\gamma_i(N)$ converge also to the
Lyapunov exponents of the matrix $T$. However, as long as one deals with
finite samples, the effective Lyapunov exponents $\gamma_i(N)$ depend on the
realization of the disorder. It is therefore convenient to average
$\gamma_i(N)$ over the ensemble of all possible realizations. In order not
to overload the notations, this average will be always understood.

\section{Scaling behaviour}

Let us first discuss the scaling behaviour of the Lyapunov exponents for
the Anderson model. In fact, the transfer matrix approach reveals a clear
analogy between the physical problem considered in the present paper and
the Anderson localization in a stripe. Indeed, the band-width $b$ plays
the same role as the strip-width $L_t$ in the sense that both define the
number of possible channels for electronic conductance. It is important to
note, however, that for the analogy to be kept as strict as possible, one
must assume that the Lyapunov exponents are measured in units of the
interaction range (or, equivalently, in mean free paths), i.e. by referring
to the lattice spacing in the Anderson problem and to $b$ in the present
case. This feature was already noticed in \cite{KPIR96}, where it was
pointed out that the Lyapunov spectra of BRM, measured in natural units,
scale as $1/b$.

One cannot straightforwardly apply to the present case the single-channel
scaling theory to infer localization properties for different disorder
amplitudes and ``transversal widths'', since there is no proper localization
length in the thermodynamic limit $b \to \infty$. In fact, while it is
conjectured that the minimum Lyapunov exponent is finite in the limit of
infinitely large stripes ($L_t \to \infty$) in the Anderson problem (in the
insulating regime), it vanishes as $1/b$ for band-random matrices, even
using the appropriate spatial scale. Let us indeed recall that the
localization length for the eigenfunctions of energy $E$ is
$l_{\infty} (E) = (2b^2/3) [1-(E^2/(8b))]$ \cite{theory}, i.e. it
diverges as $b^2$ in the whole energy range.

Another feature of the scaling behaviour that has been investigated in the
Anderson problem concerns the dependence of the Lyapunov exponents on the
sample length for fixed transversal width $L_t$. In Ref.~\cite{PZIS90} it
has been found that the scale dependence for the Anderson model has the
form
\be
       \gamma_j(N) = \gamma_j(\infty) F_j(\gamma_j(\infty)N) \quad ,
\ee
independently of the disorder amplitude. In the context of BRMs, the above
relation is somewhat trivial, since the amplitude of the disorder can be
scaled out. This is immediately realized by noticing that the elements of
the transfer matrices involve only ratios of the disorder terms (apart from
the energy term which is the only contribution that need being appropriately
rescaled), thus revealing that their absolute scale is irrelevant.

Meanwhile, a different approach has been suggested to describe scaling
properties in BRMs without leads. Such a procedure passes through the
introduction of the generalized localization lengths $l_q(N)$ \cite{CMI90}
of a generic eigenvector,
\be
\label{enle1}
l_q(N) = \exp \langle{\cal H}_q\rangle 
\ee
where ${\cal H}_q$ is defined as
\begin{equation}
\label{enle2}
{\cal H}_q = {\frac 1{1-q}} \ln P_q ;\,\,\, P_q = \sum _{n=1}^N |\psi_n|^{2q}
;\,\, \,q \geq 2 \, \,\,\,\,\,\,\
{\cal H}_1 = - \sum_n |\psi_n|^2 \ln(|\psi_n|^2)
\end{equation}
and $\psi_n$ is the $n$th component of an eigenvector of the matrix. The
average of ${\cal H}_q$ in Eq.~(\ref{enle1}) is performed over disorder and
over the eigenstates corresponding to energies within a prespecified window.
It was numerically shown in \cite{CMI90} and analytically proved
in \cite{theory} that the rescaled localization length
$l_q(N)/l_q^{GOE}(N)$, where $l_q^{GOE}$ corresponds to full random
matrices, depends only on the ratio $l_{\infty}/N $. More detailed
analytical studies \cite{theory} have revealed that the scaling behavior for
$l_q(N)$ is very close to the form
\begin{equation}
\label{enle3}
l_q^{-1}(N) = l_q^{-1}(\infty) + C_q/N \; \quad ,
\end{equation}
for $q \ne 2$, while it holds exactly for $q=2$. Notice that, in the latter
case, the localization length $l_2(N)$ is related to the inverse
participation ratio which has the simple physical meaning of the probability
for a quantum particle to return to the initial position after infinite
time. The second term in the r.h.s. of expression (\ref{enle3}) represents
the normalization factor $l_q^{GOE}(N)$: it was found both numerically and
analytically that the coefficient $C_q$ is always positive and independent
of $N$. The positiveness of $C_q$ indicates that the finite-length estimates
of the localization length converge to its asymptotic value $l_q(\infty)$
from above.

Let us finally mention that the scaling relation Eq.~(\ref{enle3}) appears
to be rather general as revealed by numerical simulations performed in
many other models like the Kicked Rotator \cite{CGIS90,I90}, 1-dimensional
Anderson, Lloyd \cite{CFGIM92}, and 1-dimensional dimer models
\cite{IKT96,VP97}.

\section{Numerical Analysis}

\subsection{The method}

As is mentioned in Sec.~II, the conductance of a finite sample can be
determined from the eigenvalues of $F^{\dagger} F$. For this reason,
here we study the convergence properties of the latter quantities
by varying the sample-size $N$. The standard technique for the
determination of the eigenvalues runs into troubles already for relatively
short samples because of numerical inaccuracies due to the small
denominators in Eq.~(\ref{eq:mmap}). Such a difficulty can be overcome by
adapting a specific algorithm for the computation of the Lyapunov exponents
of an infinite product of matrices \cite{BGGS80}. In fact, given a finite
sample of length $N$ (and the corresponding matrix $F$ defined as in
Eq.~(\ref{trvel})), one can formally construct the following infinite
product of matrices
\begin{equation}
\ldots F^{\dagger }F\ldots F^{\dagger }F\ldots F^{\dagger }F\ldots
\quad .
\label{numtec}
\end{equation}
Such a sequence can be recursively applied to $b$ independent vectors,
orthonormalizing them every single step. Accordingly, one finds $b$ Lyapunov
exponents that are nothing but the logarithms of the (real) eigenvalues of
$F^{\dagger} F$. The advantage of this procedure over the standard
diagonalizations methods is that the orthomormalization can be implemented
for all intermediate steps in the construction of $F$ (i.e. multiplication by
the transfer matrices $T$ and application of the similarity
transformations). This approach proved already its effectiveness in the
study of the Anderson problem \cite{MK93}.

In practice, the number $M$ of ``replicas'' of $F^{\dagger}F$ in
(\ref{numtec}) is finite: we have chosen $M$ so as to guarantee at least an
accuracy $10^{-4}$ for all the Lyapunov exponents, which means $M >2500$.

\subsection {Convergence to the asymptotic limit}

The main goal of this subsection is the study of the convergence of
$\gamma_i(b,N)$ towards the set of self-averaged Lyapunov exponents
${\gamma_i(b,N \rightarrow \infty)}$ as a function of the sample
length $N$ and of the bandwidth $b$.

We have already seen that BRMs have somewhat peculiar properties which make
problematic the application of the standard scaling theory to Lyapunov
exponents. A problem which, instead, makes perfectly sense also in the
context of BRM, is the attempt to combine in a single relation the dependence
of the Lyapunov exponents on the sample length $N$ and the 
``transversal'' width $b$. This is an issue
that has not yet received a clear answer in the 2- and 3-dimensional
Anderson problem \cite{PZIS90}.

The natural starting point is represented by Eq.~(\ref{enle3}), which
gives the localization length as measured directly from
the eigenfunctions of the Hamiltonian. By recalling the $b^2$-dependence
of $l_q(\infty)$, one realizes that the finite-size corrections do depend
only on a scaling parameter, namely $N/b^2$. Accordingly, one could expect
that the proper scaling relation for the Lyapunov exponents is of the type
\be
\gamma_i(b,N)/\gamma_i(b,\infty) = f(i/b , N/b^2) \quad ,
\label{scare1}
\ee
where we have added an $i/b$ dependence to account for possible differences
exhibited by the various exponents. However, a careful analysis of
our data definitely rules out such a possibility, so that one needs
to modify the above Ansatz in a more suitable manner. After many different
attempts to find the correct scaling dependence on $b$ and $N$, we have
come to the conclusion that the most convincing and yet simple scaling
hypothesis is
\be
\gamma_i(b,N)/\gamma_i(b,\infty) = f(\gamma_i(b,\infty) N b^{\alpha} )
\quad .
\label{scare2}
\ee
with $\alpha$ as some function of the ratio $i/b$.
An effective test of the scaling relation can be made after subtracting
the asymptotic value 1 from both sides of Eq.~(\ref{scare2}), i.e. by
studying the behaviour of the difference
\be
    \delta \gamma_i(b,N) \equiv 1 - \gamma_i(b,N)/\gamma_i(b,\infty) \quad .
\ee
First, it is necessary to determine the asymptotic Lyapunov exponents 
($N \to \infty$) for different bandwidths. In this limit, the ``contacts'' 
between the the ordered regions and the sample play no role as well as the 
similarity transformations involved in the definition of $F$. Accordingly, 
one can get rid of most of the technical difficulties and determine 
$\gamma_i(b,\infty)$ by resorting to the usual transfer matrix approach as 
implemented in Ref.~\cite{KPIR96}. The results reported in Fig.~1 indicate a 
convergence of the type $1/b$ in the bulk of the spectrum.

Once the asymptotic values of $\gamma_i$ have been determined, we have
plotted the finite-size correction $\delta \gamma_i$ versus the rescaled
sample length $m = \gamma_i(b,\infty) Nb^\alpha$ for different
choices of $\alpha$. In all cases, we find that $\delta \gamma_i$ is positive, 
indicating that the convergence to the asymptotic values is from below. This 
striking difference with the behaviour of the directly computed localization 
length
(see Eq.~(\ref{enle3})) is the clearest indication that the influence of the
leads and the type of contacts results in strong finite-size corrections.

The best data collapses obtained for $i/b = 0.3$, $0.5$ and $0.9$ are
reported in Fig.~2. In all cases, $f(m)$ turns out to be essentially equal
to $1+ A/m$ with the value of $A$ depending very little on $i/b$
($A \approx 1.1$). As anticipated in Eq.~(\ref{scare2}), the main
dependence on $i/b$ is contained in the exponent $\alpha$ which appears to
change linearly with $i$
\be
\alpha = i/b - 0.3
\label{linear}
\ee
Notice that the necessity to introduce a different scaling
Ansatz,  Eq.~(\ref{scare2}), in substitution of Eq.~(\ref{scare1}) is
contained precisely in the above expression of $\alpha$. In fact, only
the equality $\alpha = -1$ for $0< i/b < 1$ would reconcile the two scaling
functions. A partial justification for this result comes from the
observation that the bulk of the Lyapunov spectra scales in a different way
from the minimum value (to which Eq.~(\ref{scare1}) refers). However, this
is not enough to understand why the various parts of the Lyapunov spectrum
exhibit different convergence properties.

A clear exception to relationship (\ref{linear}) is found by analyzing
the behavior of the maximal exponent. This is not a surprise, since in
Ref.~\cite{KPIR96} it was already noticed the existence of a
``phase-transition'' in the Lyapunov spectrum occurring approximately at
$i/b=0.1$. This is illustrated in Fig.~3 where $\chi = \gamma_i i$ is
plotted versus $i/b$ revealing an incipient discontinuity in the derivative
of the spectrum. It is thus reasonable that different convergence properties
are observed above and below $i/b \approx 0.1$. Actually, we find that
the convergence is of the type $1/N$ in both cases, but the value of
$\alpha$ is -1 for the maximum exponent, as seen in Fig.~4 
(it should be recalled that
$\gamma_1$ exhibits a different scaling behaviour from that in the bulk
of the spectrum, being independent of $b$).

The behaviour of the minimal exponent is another important test, but
before discussing this case, we would like to stress that the scaling of
$\gamma_b$ as $1/b^2$ makes the numerical computations much more difficult: in
fact, it is very hard to get rid of statistical fluctuations when $b$
becomes relatively large. An indication of this difficulty is already
revealed by the comparison of the best overlaps obtained in the various
cases: increasing fluctuations are detected upon increasing $i/b$, testifying
to the importance of statistical fluctuations (this is particularly
visible in Fig.~2c, i.e. for $i/b = 0.9$). Nevertheless, one can see in
Fig.~5 that the data collapse is still not bad by assuming $\alpha = 0.7$,
i.e. the value predicted by the linear law (\ref{linear}). Consistency with
Eq.~(\ref{scare1}) would require $\alpha=0$, which gives definitely a much
worse overlap of the various curves and has to be, therefore, ruled out.
Moreover, a regression of the various curves corresponding to different
values of $b$ seems to suggest that the convergence to the asymptotic value
in this last case is slower than $1/N$, but it is not clear whether this is
an artifact due to a lack of sufficient statistics or whether it is a
finite-size ($N$) effect.

After having presented a possible unified description of the convergence
properties of ``finite-length'' Lyapunov exponents, it is worth discussing the
origin of the corrections expressed by $f$ in Eq.~(\ref{scare2}). One
reason for these corrections is the presence of the ``contacts''. Since
the ``contacts'' are less disordered compared to the bulk, one can expect
that they are characterized by a different, smaller, growth rate. The first
prediction of this argument is a negative sign for the correction, i.e.
a convergence from below, as it is indeed observed. This accounts
for the main difference found with the direct investigation of the
localization properties. Moreover, if this were the only source of
corrections, and if there were no boundary effects between the leads and the
sample, one should conclude that the relative correction must be
proportional to $b/N$ i.e. proportional to the ratio between the length of
the leads and the length of the sample. This statement is equivalent to
saying that Eq.~(\ref{scare2}) holds with $\alpha = -1$ for the maximal
Lyapunov exponent. As we have already seen, this prediction is perfectly
confirmed. Furthermore, direct numerical simulations made to compute
separately the growth rate in the contacts and in the sample do reveal that
the former contribution is half of the latter almost independently of $b$.
This proves directly the correctness of our simple conjecture.

The same argument, applied to the rest of the spectrum, would imply that
the correction is still of the order of $b/N$ which means that the $\alpha$
value in Eq.~(\ref{scare2}) is zero independently of $i/b$ (except for the
minimum). Since a strictly positive $\alpha$ is found, instead, for
$i/b>0.3$, this means that the actual correction is smaller than expected
from the above argument. We can only give a qualitative explanation
for the discrepancy: as long as $i/b$ is strictly larger than $1/b$ (in the
asymptotic limit $b \to \infty$), the correspoding growth rate $\gamma_i$ is
of the order of $1/b$ (except for the limit case of the minimum exponent) so
that the contribution to the expansion observed in the ``contacts'' is not
uncoupled to the expansion in the rest of the sample and this makes the
subdivision of the entire sample into a bulk and two ``contacts'' less
defined. Moreover, we should add that even in the absence of the leads,
finite-size corrections must be present and, as far as we know, there are
no theoretical predictions about this kind of corrections apart from the
maximal exponent \cite{PP}.

\section{Conclusions and perspectives}

We have investigated the finite-length Lyapunov spectra of symmetric Band
Random Matrices describing quasi 1-dimensional and 1-dimensional disordered
systems with long-range interactions. Our main goal was to investigate the
scaling properties of Lyapunov spectra upon changing both the bandwdith $b$
and the sample-size $N$. To our knowledge, the only example in the
literature where a similar question has been
addressed is Ref. \cite{PZIS90}, where the authors have commented about the way
to combine the scaling behaviour with the sample length and with the strip
width. However, their conjectures are not supported by numerical
analysis. Here, instead, a detailed numerical investiagation suggests that
all the convergence properties can be described in terms of a scaling
parameter $b^\eta/N$ with $\eta$ depending on $i/b$ (here, we refer to
$\eta$ instead of $\alpha$ to understand the dependence of
$\gamma_i(b,\infty)$ on $b$ as well). Such a behavior does
contrast with our expectations a priori of a $b^2/N$ dependence based
on the results of a direct analysis of the localization properties.

Even more striking is, in contrast to the 2-dimensional Anderson model, the
negative sign of the finite-size correction terms. These results can
be partly attributed to the influcence of the ``contacts'' connecting the
ordered leads with the disordered sample. However, while the $b/N$
dependence for the maximal exponent is also supported by a simple
theoretical argument, the same cannot be said for the other exponents.
Nevertheless, we wish to recall that other combinations of simple
functions provide definitely less accurate descriptions of the observed
data. On the other hand, we cannot exclude that the relative ``smallness''
of the values of $b$ and $N$ accessible to a numerical analysis, masks a
dependence different from that conjectured in Eq.~(\ref{scare2}).
We can only stress that the scaling Ansatz is chosen as by far as the
simplest one in a set of even less convincing alternatives. In other words,
our work represents an instance of the application of the Occam's razor.
Thus, some theoretical progress is needed to shed further light on this
problem. This is particularly true for the minimal exponent which is the
most delicate one to be numerically determined.

\section{Acknowledgments}

We would like to acknowledge L.~Molinari for useful discussions on
scattering problems. One of us (T.K.) would like to thank the
Istituto Nazionale di Ottica for the hospitality during the fall of 1995 and
to acknowledge the support of Grant CHRX-CT93-0107. He also wishes to
thank Prof. R.\thinspace Livi for the kind hospitality in his office
during weekends, and Prof. U. Smilansky for the stimulating environment
at the Weizmann Institute. F.M.I. acknowledges support from the INTAS grant
No. 94-2058.

\begin{figure}
\caption{Convergence of the asymptotic Lyapunov exponents 
$\gamma_i(b,\infty)$ upon increasing $b$. The three data sets refer to 
$i/b=0.3$ (plusses), $0.5$ (diamonds), and $0.7$ (circles). In all cases, 
the difference $\delta \gamma_i = \gamma_i(b,\infty)-\gamma_i(\infty,\infty)$ 
is plotted versus $b$.  The straight lines follow from a best fit: their 
slopes are always close to -1 (with a few percent deviations).}
\end{figure}

\begin{figure}
\caption{Rescaled finite-size corrections $\delta \gamma_i$ to the Lyapunov
exponents versus the rescaled length of the sample size $m$: circles 
correspond to $b=10$, diamonds to $b=20$ and plusses to $b=40$;
(a), (b), and (c) correspond to $i/b = 0.3$, $0.5$ and $0.9$, respectively.
The straight lines are the best fits. The deviations of the resulting
slope from -1 is approximately 3\%  in all cases.}
\end{figure}

\begin{figure}
\caption{Lyapunov spectrum as determined for $b=10$ (circles), 20 (squares), 
30 (diamonds), and 40 (triangles).}
\end{figure}

\begin{figure}
\caption{Finite-size corrections of the maximum Lyapunov exponent, using the
same representation as in Fig.~2. The slope of the straight line is -1.}
\end{figure}

\begin{figure}
\caption{Finite-size corrections for the minimum Lyapunov exponent, using the
same representation as in Fig.~2. The straight line with slope -1 is drawn
for reference.}
\end{figure}

\end{document}